\documentclass[12pt,preprint]{aastex}

\newcommand{\sgra}    {\mbox{\rm\,Sgr~A$^*$}}

\slugcomment{To appear in ApJ, 20 February 2003 issue}

\begin{document}

\title{A New X-Ray Flare from the Galactic Nucleus
Detected with the XMM-Newton Photon Imaging Cameras}

\author{A. Goldwurm, E. Brion\altaffilmark{1}, 
P. Goldoni, P. Ferrando, 
F. Daigne, A. Decourchelle,
}
\affil{
Service d'Astrophysique, DAPNIA/DSM/CEA, CE-Saclay, F-91191 Gif-Sur-Yvette, France}
\author{R.S.~Warwick,}
\affil{
Department of Physics and Astronomy, University of Leicester, Leicester LE1 7RH, UK}
\author{P. Predehl}
\affil{
Max-Planck Institut f\"{u}r Extraterrestrische Physik, Postfach 1312, 85741 Garching, Germany}

\altaffiltext{1}{also Astronomical Observatory and University ``Louis Pasteur'',
Strasbourg, France}

\begin{abstract}

\sgra, the compact radio source, believed to be the counterpart 
of the massive black hole at the galactic nucleus, 
was observed to undergo rapid and intense flaring activity in X-rays 
with Chandra in October 2000. 
We report here the detection with XMM-Newton EPIC cameras 
of the early phase of a similar X-ray flare from this source, 
which occurred on September 4, 2001.
The source 2-10~keV luminosity increased by a factor $\approx$~20 
to reach a level of 4~10$^{34}$~erg~s$^{-1}$ in a time interval 
of about 900~s, just before the end of the observation.
The data indicate that the source spectrum was hard 
during the flare.
This XMM-Newton observation confirms the results obtained by Chandra and 
suggests that, in \sgra, rapid and intense X-ray flaring is not a rare event.
This can constrain the emission mechanism models proposed for this source, 
and also implies that the crucial multiwavelength observation 
programs planned to explore the behaviour of the radio/sub-mm and
hard X-ray/gamma-ray emissions during the X-ray flares, 
have a good chance of success.
\end{abstract}


\clearpage
\newpage

\section{Introduction}

The bright, compact and variable radio source 
\sgra\ is believed to be the radiative counterpart of the
2.6~10$^{6}$~M$_{\odot}$ black hole which governs the dynamics 
of the central pc of our Galaxy \cite{mefa01}.
The compelling evidence for the presence of a dark mass 
concentration at the Galactic Center \cite{gen97,ghe00}, 
which implies the presence of a massive black hole, 
contrasts remarkably with the weak high-energy activity
of such an extreme object.
In spite of the fact that some amount of material,
either provided by stellar winds from a close stellar cluster 
or by the hot surrounding medium, is probably feeding 
a moderate/low level of accretion, 
the total bolometric luminosity of the source amounts to less 
than 10$^{-6}$ of the estimated accretion power 
\cite{mefa01,gol01}.

This has motivated the development of several black hole accretion 
flow models with low radiative efficiency, some of which have  
also been applied to binary systems, low luminosity nuclei 
of external galaxies and low luminosity active galactic nuclei.
These models include spherical Bondi accretion in conditions
of magnetic field sub-equipartition with a very small 
Keplerian disk located within the inner 50 Schwarzschild radii (R$_S$), 
large hot two-temperature accretion disks dominated by advection (ADAF) 
or non-thermal emission from the base of a jet of relativistic electrons
and pairs, and some other variants or combination of the above 
(see the review by Melia $\&$ Falcke (2001)).
However any such model still predicts some level of X-ray emission 
from \sgra\ and determining the properties of such emission
would constrain the theories of accretion and outflows in the 
massive black holes and in general in compact objects.

The search for high energy emission from \sgra\ with focussing X-ray 
telescopes dates back to the end of the '70s \cite{pre94,bag01a,gol01}, 
but has recently come to a turning point with the remarkable observations 
made with the Chandra X-ray Observatory in 1999 and in 2000. 
Baganoff et al. (2001a) first reported the detailed 0.5$''$ 
resolution images obtained
with Chandra in the range 0.5-7~keV, which allowed, finally, the detection 
of weak X-ray emission from the radio source. 
The derived luminosity in the 2-10~keV band was 2~10$^{33}$~erg~s$^{-1}$,
for a distance of 8~kpc,
and the measured spectrum was steep, with power law photon index of 2.7.
Marginal evidence that the source is extended on a 1$''$ scale 
was also reported, but at low significance level.
Then, in October 2000, the same source was seen to flare up by
a factor of $\approx$~45 in a few hours \cite{bag01b}.
The luminosity increased from a quiescent level similar to the one
measured in 1999 to a value of 10$^{35}$~erg~s$^{-1}$. 
The flare lasted a total of 10~ks but the shortest variation 
took place in about 600~s, implying activity on length
scales of $\approx$~20~R$_S$, 
for the above quoted mass of the galactic center black hole.
Evidence of spectral hardening during the flare was also reported by 
the authors who determined a source power law photon index during
the event of 1.3 ($\pm$~0.55), 
significantly flatter than observed during the quiescent state.
These results constrain models of the accretion flow and radiation
mechanism for \sgra. A confirmation of the Chandra results and in particular
a better determination of the flaring properties of the source are 
therefore crucial for the modeling of the physics of the 
Galactic Center and in general for the theories of accretion 
in black hole systems.

XMM-Newton, the other large X-ray observatory presently in operation, 
features three large area X-ray telescopes 
coupled to three CCD photon imaging cameras (EPIC) operating 
in the 0.1-15 keV range and to two reflection grating spectrometers 
(RGS) working in the 0.1-2.5 keV band \cite{jan01}. 
Although its angular resolution (6$''$ FWHM) is insufficient for
properly resolving \sgra\ in quiescence,
the high sensitivity and wide spectral range of XMM-Newton allow deep 
studies of the X-ray emission of such a complex and crowded 
region like the Galactic Center.
Indeed an intense flare such as the one seen by Chandra 
can be easily detected
with XMM-Newton thanks to its large effective area, and its timing
and spectral properties can be studied.

The Galactic Center region is one of the priority targets of the 
XMM-Newton mission and was included in the guaranteed time program.
Visibility constraints and solar flare events have however
delayed the monitoring of the very center of our Galaxy.
A complete pointed observation was finally performed in fall 2001 and 
in this letter we report the detection with XMM-Newton of another X-ray 
flare from \sgra\ which occurred at this time.

\section{Observations and Results}

XMM-Newton was pointed towards the galactic nucleus for about 26~ks 
on 4$^{th}$ September 2001. This observation was part of a large 
survey program of about 10 overlapping XMM-Newton pointings planned to map 
the Galactic Plane within 1$^{\circ}$ from the Center. 
Preliminary mosaiced images of the region have been presented 
by Warwick (2002) who showed that the region is complex and dominated by 
diffuse emission and some point-like and extended bright sources.
We report here results obtained with the 
EPIC cameras of XMM-Newton during the observation of the survey which 
was directly pointed towards \sgra\ (observation GC6). 
The purpose of our analysis was to search for X-ray variability 
from this source of the type observed during fall 2000 
with the Chandra telescope.

The observation with the EPIC MOS cameras \cite{tur01}
started at 01:27:08 (UT) and lasted 26127~s 
while the PN \cite{str01} was activated 4109~s after the MOS 
for a total exposure of 21748~s.
The EPIC cameras were used in standard {\it Full Frame} imaging mode
with the medium filter while the RGS was used in {\it spectral+Q} 
mode and the OM was blocked.
Data reduction was performed using the XMM-Newton Science Analysis 
Software (SAS) 
package (version 5.3), using standard choices to select events
for both MOS and PN cameras, namely the 0 to 12 patterns for the MOS
and both the single and double events for the PN.
Inspection of the integrated count rate of the CCDs versus time revealed 
that the observation was slightly perturbed by weak soft proton flares. 
The CCD light curves show several peaks all along the
observation and in particular at the beginning and at the end of it. 
The average combined count rate in the central CCD of the two MOS cameras  
was about 4.50 cts~s$^{-1}$, while peaks due to the proton flares 
reached maximum values of 21~cts~s$^{-1}$ for short 
time intervals. 
The counts arising from the flares are distributed fairly uniformly 
across the CCDs (i.e. with only a modest vignetting effect)
and indeed derived images did not show particular features due to 
these background events.
The image recorded in the central CCD (11$'$~$\times$~11$'$ for the MOS) 
is dominated by the diffuse emission of the Sgr A East region 
which is thermal in origin and rich in emission lines.
Proper analysis of this component involves modelling of the variable
background and composition of different images of the survey. 
The work is in progress and results will be reported elsewhere.
Here we rather concentrate on the analysis of the central point-source,
\sgra, and in particular on the search for variability of 
its X-ray emission.

In order to optimize the signal to noise of the central source 
in the presence of the strong diffuse emission and considering 
the typical width of the XMM-Newton point spread function 
(15$''$ half power diameter),
we extracted and analyzed light curves of events collected within 
a 10$''$ radius region centered on \sgra.
As shown in Fig.~1, the 2-10 keV count rate 
from the combined MOS~1 and MOS~2 events selected in this way,
is quite stable around an average value of 0.08~cts~s$^{-1}$ till the last
900~s of the observation. 
Then the count rate gradually increases to reach a value of about a factor 
3 higher in the last bin. 
The integrated count rate in the last 900~s reaches 7~$\sigma$ over 
the average value measured before the flare and the detected variation 
has a very low probability to be a statistical fluctuation.

A similar light curve, from a 9 times larger region of the central CCD,
far from the source but including a bright part of the diffuse emission, 
is also reported for comparison in Fig.~1 (scaled for clarity),
and does not show any evidence of such an increase in the counting rate. 
The same trend is observed in the counts extracted from the PN 
(see Fig.~1 right). Since the
PN camera stopped observing about 250~s before the MOS, the last part 
of the flare is not visible. However the increase in the last (PN) 600~s
is also highly significant (4.3~$\sigma$) and again it is not detected in 
counts extracted from a region away from the source.
Some proton flares occurred in the last part of the observation,
and they increase the total CCD 2-10 keV count rate 
of about a factor 1.4 in the last 1000~s. However a detailed light curve of 
these events show that they occur with a different time behavior
than the source flare and, because of their uniform spatial distribution, 
they give rise to only a $\approx$~1.5~$\%$ increase in the measured count rate 
within the 10$''$ radius region during the last 900~s of the observation.

In order to check that the flare actually originated in \sgra, 
we constructed images using MOS events selected in different 
periods of the observation.
In Fig.~2 we report an image of the region around the nucleus
integrated during the 1000~s before the flare and a similar image 
integrated during the last 1000~s and fully including the source flare.
The basic image pixels were rebinned by a factor 5, 
giving image pixels sizes of 5.5$''$ width. 
The brightening we detected in the light curves is clearly due to 
the brightening of a central source with the counts in the central pixel 
of the image which increase from 10 before the flare to 40 during the event.
The general level of the image during the flare is a factor 1.4 times higher 
than before the flare (for pixels $>$ 20$''$ from \sgra) 
due to the proton flaring events.
This difference corresponds to the uniform increase in counts due to the 
presence of the proton flares in the last part of the observation.

We also used the SAS procedure ({\it eboxdetect}) 
for source detection to determine the location of the excess. 
On the 2-10 keV MOS 1 and MOS 2 image of the last 1000~s, rebinned 
to have pixel size of 4$''$, we obtained the centroid of the source at 
R.A. (2000) = 17$^h$~45$^m$~39.99$^s$ 
~Dec (2000) = -29$^{\circ}$~00$'$~26.7$''$,
with a pure statistical error of 0.4$''$ (90$\%$ confidence level 
in one parameter) in each direction.
To check that the attitude reconstruction does not suffer from large
systematic errors, we checked the full low energy (0.5 - 4 keV) MOS 
images for X-ray sources with stellar counterparts.
In the full field of view of MOS~1
we identified 6 point-like sources with stars of known position, 
we determined the offsets between our derived positions and optical 
positions, and computed the average and rms values.
We obtained an average offset of -0.03$''$ in right ascension
and 0.20$''$ in declination, with rms of 0.29$''$ (R.A.) 
and 1.50$''$ (in Dec.).
This implies that the absolute accuracy of location for the central CCD 
is not worse than the residual systematic uncertainties in the 
XMM-Newton focal plane, estimated to be 1.5$''$ \cite{kir02}.
The derived flare position is therefore compatible with \sgra\ 
radio location \cite{zad99}, since it is offset from the latter 
by only 1.5$''$ (of which 1.4$''$ in Dec.), i.e. within uncertainties.

The detailed study of Chandra data carried out by 
Baganoff et al. (2001a) showed 
that a number of other X-ray point sources are present in the vicinity
of the galactic nucleus, the nearest of which is 
associated with the infrared and radio object called IRS~13 at 
an angular distance of 4$''$.
The error box we derived with XMM-Newton excludes the possibility that 
the observed flare is due to IRS~13 (offset $>$ 4$''$) 
and also is not consistent with another region of excess counts in the
Chandra images (offset $\approx$ 3$''$), not fully recognized as a 
source but tentatively identified with IRS~16SW by Baganoff et al. (2001a).
Of course this does not completely rule out the possibility 
that another X-ray source in this crowded region of the sky 
may be the origin of the event.
However due to the close resemblance of the event with the Chandra flare,
which was more precisely located on \sgra, and considering that 
generally X-ray binaries do not show such large variations on such short 
timescales, it seems unlikely that another
X-ray source in the central star cluster is responsible for this flare.
We conclude that the flare detected by XMM-Newton is associated with \sgra.

A first spectral analysis of the flaring event was performed by computing 
a simple hardness ratio (H/S), 
defined as the ratio between the measured counts (including background) 
in the hard band 4.5-10 keV, and those in the soft band 2-4.5 keV. 
Using events collected by both MOS and PN cameras within 
10$''$ from \sgra\ we computed the variation of the hardness ratio during
the flare compared to the average value measured before the flare.
The measure was performed separately for MOS (flare during last 900 s) 
and PN (last 650 s) events and then we computed the weighted average.
The hardness ratio increased by 0.32 $\pm$ 0.127 
during the event with respect to the value before the flare.
Though the count rate hardening has a modest statistical significance 
of 2.5~$\sigma$, 
it is fairly consistent with the trend observed with Chandra
for the \sgra\ flare.
Unlike Baganoff et al. (2001b) who observed the flare at
low energy to follow of few hundred seconds the flare at high energy, 
we have not revealed any 
significant lag between the light curve of the soft energy band 
and the one of the high energy band.

To derive a spectrum of the source during the flare state
we have to model the emission which is not due to \sgra. 
Detailed analysis of Chandra data showed that only $\approx$~10$\%$ 
of the 2-10 keV emission measured within 10$''$ from the galactic 
nucleus is due to \sgra\ in quiescence, 
the rest is mainly due to the local diffuse component (60$\%$),
to the contribution of the six other point sources seen by Chandra 
(20$\%$) and to more diffuse Galactic emission together with 
the instrumental background (10$\%$).
To model the thermal component Baganoff et al. (2001a) used the 
Raymond $\&$ Smith hot gas spectral model with twice solar abundance
\cite{ray77}. The contribution from six point sources in this region was
modeled with a power law of photon index 
$\approx$~2.5, while the contribution from \sgra\ was best fitted 
by a power law of index 2.7 and N$_H$ = 9.8 10$^{22}$~cm$^{-2}$. 
The Chandra derived best fit parameters (see Table 3 of Baganoff et al. 2001a) 
were used to evaluate, by convolving the model with XMM-Newton responses,
the expected count rates in the 2-10 keV band before the flare. 
We obtained values very close to the measured ones
(demonstrating that the contribution of the instrumental background 
is not a major influence for this region of high X-ray surface brightness) 
and therefore we adopted a similar model to fit XMM-Newton data.

We extracted MOS and PN count spectra from the 10$''$ radius
circular region centered on \sgra\ before the flare and
during the flare (last 900~s for MOS and last 700~s for PN).
For the spectra before the flare, rebinned 
to have 30 counts per bin, we used the above model
for which we fitted all parameters simultaneously (tied) on MOS 
and PN data, apart from the \sgra\ parameters 
which were kept frozen to Chandra best fit values (with the norm
reduced by factor 0.6 to account for the encircled energy loss).
Normalizations were left to vary untied between MOS and PN
data to allow adjustment to different instrument background 
and residual normalization differences not accounted for by the 
responses, and we obtained the best fit parameters reported in Table~1
(left column). The MOS spectrum, compared to the best
fit model, is reported in Fig.~3.
Allowing the normalization of the \sgra\ power-law to vary freely in 
the fitting 
(with fixed slope at 2.7 and fixed N$_H$ at 9.8 10$^{22}$~cm$^{-2}$) 
we obtained a very low normalization value, 
indicating that this component, if present, cannot be disentangled
from the first power-law supposed to describe point sources 
and other residual background components.
With a reduced $\chi^{2}$ of 1.35 for 121 dof, the fit is acceptable.
The best fit parameters match well the values obtained by Chandra,
in particular the gas temperature kT of 1.3~keV and its column density
are fully compatible with Chandra results,
while the power law spectrum for the point sources is slightly
flatter and needs a rather higher N$_H$.
In the procedure we adopted, this power-law describes also the background 
and we do not expect it to fully represent the point sources contribution.

We then fixed these parameters, let free the \sgra\ power-law slope
and normalization and fit the spectrum extracted during the flare.
The fit was performed simultaneously on MOS and PN data of the flare,
rebinned to have 10 counts per bin, and fixing the column 
density to the value measured by Chandra. 
The power-law normalizations for MOS and PN data were left untied 
since a different portion of the flare was seen by MOS and PN 
cameras and we expect different average intensity.
We obtained the parameters reported in Table~1 (right column).
The MOS count spectrum during the flare, and its best fit model, 
is shown in Fig.~3, compared to the MOS spectrum before the flare.
The power-law photon index of the flaring source is
0.9 $\pm$ 0.5 (error at 1 $\sigma$ for one
interesting parameter) which is significantly harder than
the spectrum measured with Chandra during the quiescent state 
(2.7 $\pm$  1.0),
and rather compatible, within uncertainties, to the
index measured during the 2000 October flare.
By allowing the column density to vary freely during the fit we obtained 
a lower value for \sgra\ (N$_H$ = 6.4 $^{+4.0}_{-3.3}$ 10$^{22}$ cm$^{-2}$) 
and an even harder slope. This value is consistent with the value 
determined by Chandra and which we use to fix the column density
in the fits.

Similar parameters were obtained by simply using the spectra extracted
before the flare as background components for the flare spectra.
After subtraction of the non-flaring count spectrum, the flaring 
spectrum was fitted with a simple absorbed power law with N$_H$ 
fixed to the Chandra measured value and leaving untied the MOS and PN 
normalizations. Results are reported in Table 2, left column. 
We obtained a photon index of 0.68 $^{+0.53}_{-0.60}$
compatible with the estimate reported in Table 1.
This procedure subtracts from the flare spectrum the 
non flaring component of \sgra\
and therefore assumes that the quiescent emission from \sgra\
is negligible. 
This is an acceptable approximation since, if the emission level of 
\sgra\ is comparable to the one observed by Chandra in 1999, 
it is expected to contribute by only $\approx$~5$\%$ 
to the counts of the flare spectrum. 
On the other hand this procedure allows to subtract
the diffuse emission present in the region of the spectral extraction 
and the instrumental background in a model-independent way.
In any case the above results and derived fluxes, 
are equivalent, within errors, to those obtained adding 
a power-law component to the non-variable emission model.
To directly compare our results with the spectra 
of the \sgra\ flare observed with Chandra,
we also fitted the above spectra with an absorbed power law model 
modified by the effect of dust scattering \cite{pre95}.
We fixed the parameter of the dust scattering model, the visual extinction A$_V$,
to the Galactic Center canonical value of 30 magnitudes and the column density 
for absorption to N$_H$~=~5.3~10$^{22}$~cm$^{-2}$ as found by Baganoff et al. 
(2001b). We then fitted the model on the MOS and PN spectra during the flare
using the spectra before the flare as background component. 
The best fit is found for an even harder power law slope of 0.31 (see Table 2, right column).
Letting the N$_H$ to vary freely, again we find a best fit value
for the column density around 5~--~6~10$^{22}$~cm$^{-2}$.

To compute the observed flux and luminosity we used the normalization value
derived from the MOS data, since the MOS observed a bigger fraction
of the flare, and we corrected for the fraction
of encircled energy at a distance of 10$''$ (60$\%$).
We note that we did not corrected for the energy dependence of the 
encircled energy, which will tend to slightly harden the spectrum, however 
the statistical errors is by far larger than the systematic 
bias induced by this procedure.
The measured absorbed source flux in the 2-10 keV band is then of
(3.3 $\pm$ 0.6)~10$^{-12}$ ph~cm$^{-2}$~s$^{-1}$, 
equivalent to a 2-10~keV luminosity at 8~kpc 
of (3.8 $\pm$ 0.7) 10$^{34}$ erg~s$^{-1}$ 
(1~$\sigma$ errors computed by fixing all other parameters 
but the \sgra\ power-law normalization at the best fit values 
listed in the right column of Table 1).
This is the average value in the last 900~s but the last light
curve 180~s bin was about a factor 1.4 higher, thus the luminosity
reached a value of 5.4~10$^{34}$ erg~s$^{-1}$.

These numbers are subject to large errors due to the low
statistics available. But the general result which emerges is that 
the flare we detected presents a harder spectrum than the one
measured with Chandra for \sgra\ during the quiescent period.
The measured slope is even harder than typically found in X-ray spectra 
of AGN, however, considering the large uncertainties, this result
is not compelling.

\section{Discussion}

The XMM-Newton discovery of a new X-ray flare of \sgra\ in September 2001
confirms the results obtained in the earlier Chandra observations.
XMM-Newton observed only the first part of the flare, 
but the recorded event is fully compatible in intensity and time 
scale with the early phase of the flare seen by Chandra.

The count rate within 10$''$ from \sgra\ 
increased in 900~s by a factor 3, but if attributed
to \sgra\, it implies that this source brightened by a factor 
about 20-30, which is compatible with the increase in the first 
1000~s of the flare observed by Chandra. We have not detected 
the maximum in the flare rise. 
Therefore we cannot strictly apply the travel light argument to estimate 
the size of the emitting region.
If we assume that the flare duration (900~s) we observed is the shortest 
time scale of variation of the present event, 
it corresponds to a size of about 30~R$_S$. 
This limit is a factor 1.5 larger than the shortest scale
estimated with the Chandra data and does not constrain further 
the geometry of the region.

On the other hand the detection of another such a flare indicates that 
the event is not rare. The total reported observation time with Chandra
amounts to $\approx$~75~ks. Considering the XMM-Newton 26~ks exposure, 
the duty cycle of such event is 0.11 (= 11~ks / 101~ks), 
but it would increase to 0.18 (= 20~ks / 110~ks) if we assume that 
the flare we detected for only 1000~s would last for 10~ks.
Though not much different than the value determined with Chandra, 
this estimate of the active time fraction of the source
is now based on 2 events and it is therefore more significant.

The radio source on the other hand has been observed many times 
and the detected flux variability has never exceeded a factor 2 
\cite{zha01}.
This implies that it is unlikely that radio or sub-mm emission
present a comparable increase in flux.
If this is confirmed the flare may not be due to a change in the accretion 
rate, since this variation would lead, at least in models which
attribute the bulk of X-ray emission to self synchrotron Compton 
emission, to a comparable increase of radio and sub-mm radiation 
\cite{mar01}.

The X-ray flare from \sgra\ cannot be explained by pure 
ADAF models (Narayan et al. 1998) as in these models the emission 
is due to thermal bremsstrahlung
from the whole accretion flow and arises from an extended region 
(between 10$^3$ - 10$^5$ R$_S$) which cannot account for such 
rapid variability. 
Models which predict emission from the innermost regions
near the black hole involve a mechanism acting either at the base 
of a jet of relativistic particles \cite{mar01}
or in the hot Keplerian flow present within the circularization radius
of a spherical flow \cite{mel01,liu02}.
In both cases a magnetic field is present in the flow and 
the sub-mm radiation is attributed to optically thin 
synchrotron emission from the inner region, while the X-rays are produced 
by the synchrotron self-Compton (SSC) mechanism whereby radio to mm 
photons are boosted to X-ray energies by the same relativistic 
or subrelativistic electrons that are producing the synchrotron radiation.
Large flux variations can be produced by a change in accretion rate 
or, in the jet model, by additional heating of the electrons caused 
for example by magnetic reconnection.
The second mechanism would increase (and harden) the X-ray flux 
without significantly increasing the radio and sub-mm part of 
the spectrum and therefore it could be more compatible with the lower
amplitude of radio changes compared to X-rays \cite{mar01}.

However even emission from a circularized flow can provide
low or anti correlation of the radio emission with the X-rays
if the radiation mechanism for the X-rays 
is bremsstrahlung rather than SSC \cite{liu02}.
The sub-mm and far IR domain on the other hand
would in this case show a large correlated increase, 
but at these frequencies the measurements have not been frequent 
enough to settle the issue.
Though the exact modelling of radiation process depends 
on viscosity behavior and other uncertain details, 
the observed hardening of the spectrum during the flare indeed
favours the bremsstrahlung emission mechanism in this model rather 
than the SSC one \cite{liu02}.
More compelling constraints on the models will be set when 
simultaneous observations in radio/sub-mm and X-ray wavelengths
of such a flare are obtained.

Correlated radio and X-ray observations are indeed crucial because,
althought the amplitude of the radio variability is low compared
to the event recorded in X-rays, an intriguing correlation
seems to be present between the X-ray flares and the rise of the 
radio emission.
Indeed Zhao et al. (2001), using Very Large Array (VLA) data collected 
over two decades, detected a periodicity in the \sgra\ radio
variability, with a 106 days cycle and a characteristic timescale 
of 25 days. Baganoff et al. (2001b) already remarked that
the October 2000 X-ray flare occured at a radio-cycle phase 
corresponding to the beginning of the radio peak.
We have computed the 106 days radio cycle phase of the X-ray flare that 
we detected with XMM-Newton and found that it differs by only 6 days 
from the phase of the flare detected with Chandra.
The flare occurred at the day 64 in the light curve of Fig.~3 of 
Zhao et al. (2001), while the Chandra flare took place at phase 70 day
and the 1.3~cm radio peak rise extends roughly from day 55 to day 75.
Even though the light curve radio peak is wide and several other 
structures are present,
both X-ray flares detected till now are very close in phase and take 
place during the rising part of the main radio flare.
We have also compared the time of the flare to a recent radio light curve
of \sgra\ obtained at 1.3 cm and 2 cm with the VLA between March  
and November 2001 \cite{yua02}. The X-Ray flare 
occurred 1-2 days after a local maximum of the curve, but no radio data
points are reported for the day when our XMM-Newton observation
took place.

It will be also important to study the shape of the flare spectrum at 
energies higher than 10 keV to fully understand the radiation mechanism 
producing the high energy tail.
In particular by measuring the high energy cut-off of the spectrum  
one could determine the electron temperature for a thermal emission 
or the Lorentz factor for non-thermal processes.
We estimated that such a flare should be marginally visible in the 
range 10-60 keV
with the low energy instruments onboard the new gamma-ray mission 
INTEGRAL, to be launched in October 2002, 
if the spectrum extends to these energies with the slope observed 
with Chandra and XMM-Newton.

Our more secure estimation of the duty cycle of the flares
shows that multiwavelength observations of \sgra\ which involve
XMM-Newton or Chandra will have good chance of observing 
an X-ray flare provided the simultaneous coverage 
is of the order of 100~ks.

\section*{Acknowledgments}

Based on observations with \textit{XMM-Newton}, an ESA science mission 
with instruments and contributions funded by ESA member states and the USA 
(NASA).
This observation was performed as part of the \textit{XMM-Newton} 
guaranteed time program of the XMM-EPIC team.
We wish to thank all \textit{XMM-Newton} staff involved in the
realization and operation of the mission.
F.D. acknowledges financial support from a postdoctoral fellowship 
from the French Spatial Agency (CNES).
We thank the referee Frederick Baganoff for his stimulating comments 
and suggestions. We also thank Ruby Krishnaswamy for 
help with installation of data analysis software.


\clearpage
\newpage
\begin{deluxetable}{llcc}
\tabletypesize{\footnotesize}
\tablewidth{0pt}
\tablecolumns{4}
\tablecaption{Spectral Fit to X-ray Emission from within 10$''$ from \sgra\ }
\tablehead{
\colhead{} &
\colhead{} &
\colhead{Spectrum} &
\colhead{Spectrum} \\
\colhead{} &
\colhead{} &
\colhead{Before the Flare\tablenotemark{a}} &
\colhead{During the Flare\tablenotemark{b}}
}
\startdata
\multicolumn{4}{l}{Optically Thin Thermal Plasma\tablenotemark{c}}\\
& $\rm N_H$ [$10^{22}$ cm$^{-2}$] & 11.1~$^{+0.6}_{-0.6}$ & 11.1 \\
& kT [keV] & 1.31~$^{+0.08} _{-0.07}$ & 1.31 \\
& Norm MOS\ [$ 10^{-3}$ cm$^{-5}$]\tablenotemark{d} & 4.6~$^{+1.0} _{-1.0}$ & 4.6  \\
& Norm PN\ [$ 10^{-3}$ cm$^{-5}$]\tablenotemark{d} & 3.6~$^{+0.9} _{-0.7}$ & 3.6  \\
\multicolumn{4}{l}{Power-law for Point Sources } \\
& $\rm N_H$ [$10^{22}$ cm$^{-2}$] ~~~~~~~~~ & 21.2~$^{+6.2} _{-4.7}$ & 21.2 \\
& Photon Index & 2.1~$^{+0.3} _{-0.2}$ & 2.1 \\
& Norm MOS [$10^{-4}$ ph cm$^{-2}$ s$^{-1}$ keV$^{-1}$]\tablenotemark{e}\  & 6.8~$^{+6.0} _{-3.8}$ & 6.8 \\
& Norm PN [$10^{-4}$ ph cm$^{-2}$ s$^{-1}$ keV$^{-1}$]\tablenotemark{e} \  & 6.3~$^{+5.5} _{-3.5}$ & 6.3 \\
\multicolumn{4}{l}{Power-law for \sgra\ } \\
& $\rm N_H$ [$10^{22}$ cm$^{-2}$] & 9.8 & 9.8 \\
& Photon Index & 2.7 & 0.9~$^{+0.5} _{-0.5}$ \\
& Norm MOS [$10^{-4}$ ph cm$^{-2}$ s$^{-1}$ keV$^{-1}$]\tablenotemark{e}\ & 2.1 & 2.0~$^{+2.3} _{-1.3}$ \\
& Norm PN [$10^{-4}$ ph cm$^{-2}$ s$^{-1}$ keV$^{-1}$]\tablenotemark{e} \ & 2.1 & 0.6~$^{+0.9} _{-0.4}$ \\
\multicolumn{4}{l}{Goodness of fit} \\
& $\rm \chi^2_{\nu}~(d.o.f.)$ & 1.35 (121) & 0.90 (27) \\
\enddata
\tablenotetext{a}{Spectrum from within 10\arcsec\ of
\sgra\ integrated in the period before the flare.}
\tablenotetext{b}{Spectrum from within 10\arcsec\ of
\sgra\ integrated in the period during the flare.}
\tablenotetext{c}{Raymond $\&$ Smith (1977) model with
twice solar elemental abundances.}
\tablenotetext{d}{Normalization in units of $10^{-14} \int n_{\rm e}
n_{\rm i} dV / 4\pi D^2$, where $n_{\rm e}$ and $n_{\rm i}$ are the
electron and ion densities (cm$^{-3}$) and $D$ is the distance to the
source (cm).}
\tablenotetext{e}{Flux density at 1~keV.}
\tablecomments{Parameters without errors are fixed. 
Errors are at 68.3$\%$ confidence interval 
for one interesting parameter.}
\end{deluxetable}

\clearpage
\newpage
\begin{deluxetable}{lcc}
\tabletypesize{\footnotesize}
\tablewidth{0pt}
\tablecolumns{3}
\tablecaption{Spectral Fit to X-ray Emission from within 10$''$
   from \sgra\ during the Flare}
\tablehead{Power-law Model &
\colhead{} &
\colhead{} \\
\colhead{} &
\colhead{No Dust Scattering} &
\colhead{Dust Scattering\tablenotemark{a}}
}
\startdata
$\rm N_H$ [$10^{22}$ cm$^{-2}$] & 9.8 & 5.3 \\
Photon Index & 0.7~$^{+0.5} _{-0.6}$ & 0.3~$^{+0.6} _{-0.4}$ \\
Norm MOS [$10^{-4}$ ph cm$^{-2}$ s$^{-1}$ keV$^{-1}$]\tablenotemark{b}\ & 1.3~$^{+2.0} _{-1.3}$ & 0.7~$^{+1.0} _{-0.4}$ \\
Norm PN [$10^{-4}$ ph cm$^{-2}$ s$^{-1}$ keV$^{-1}$]\tablenotemark{b} \ & 0.3~$^{+0.6} _{-0.3}$ & 0.2~$^{+0.3} _{-0.1}$ \\
$\rm \chi^2_{\nu}~(d.o.f.)$ & 0.98 (20) & 0.95 (20) \\
\enddata
\tablenotetext{a}{Scattering computed for fixed value of A$_{V}$~=~30.}
\tablenotetext{b}{Flux density at 1~keV.}
\tablecomments{Parameters without errors are fixed. 
Errors are at 68.3$\%$ confidence interval 
for one interesting parameter.}
\end{deluxetable}

\clearpage
\newpage
\begin{figure}
\plottwo{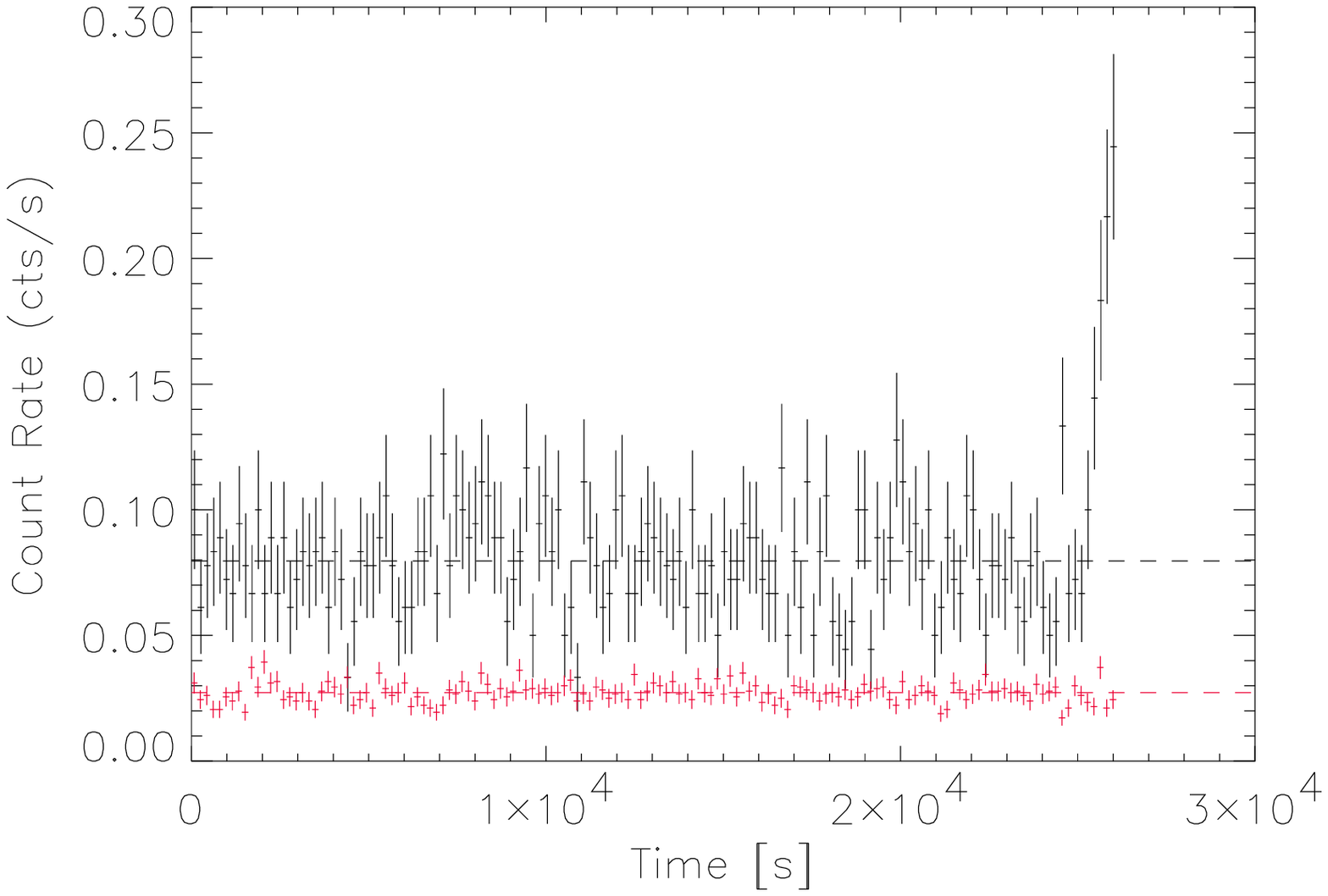}{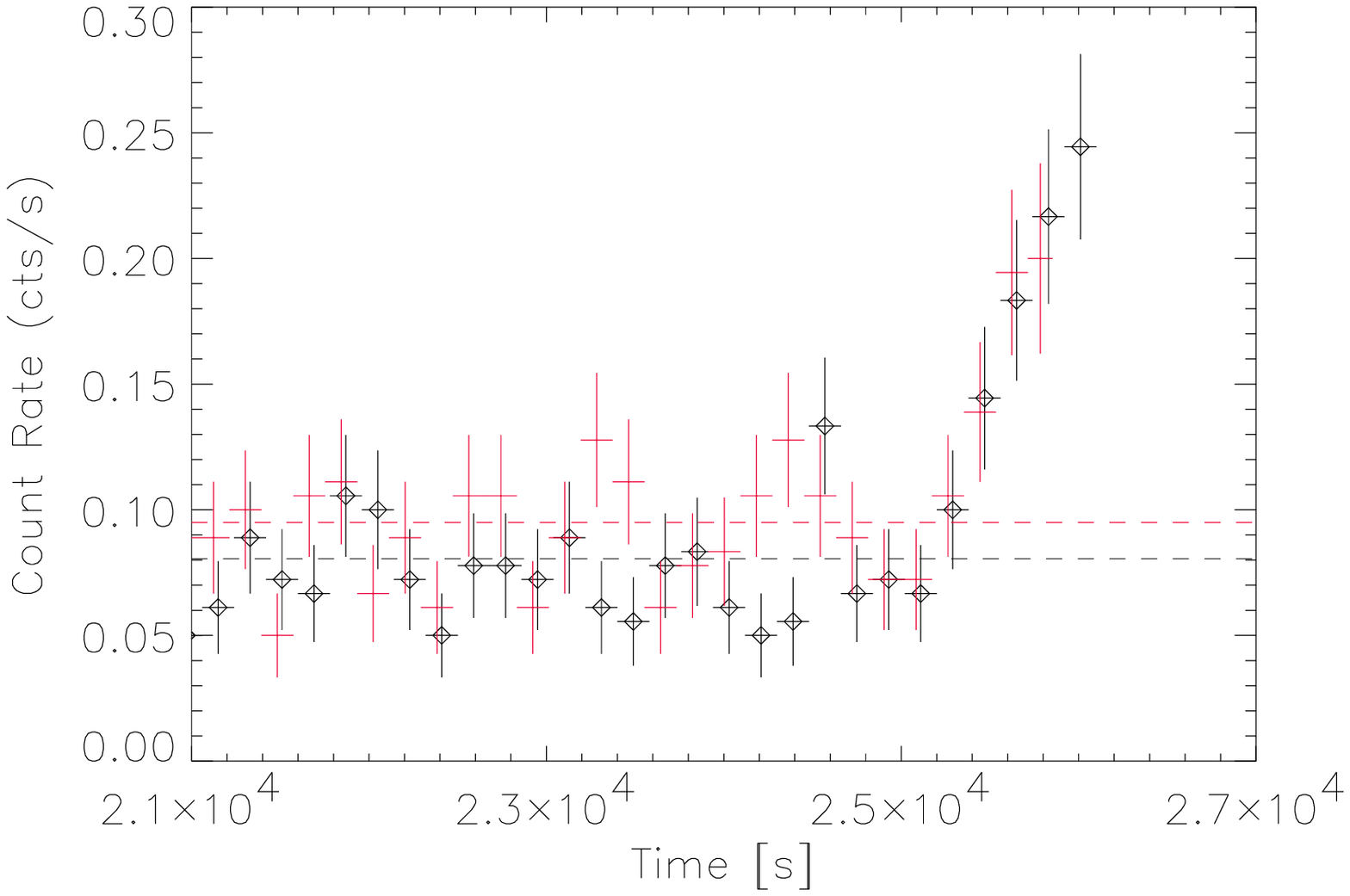}
\caption{Left: Count rate, sampled in bins of 180~s, collected with both MOS 
cameras from a region within 10$''$ from \sgra\ in the range 2-10 keV
(black upper curve). An equivalent light curve collected from a 30$''$ radius
region centered about 1$'$ East of \sgra\ and rescaled by a factor 0.1
for clarity, is shown for comparison (red lower curve).
Dashed lines indicate the average value computed before the flare.
Right : a zoom of the \sgra\ MOS light curve (black circles) around 
the period of the flare compared to a similar light curve (count rate
within 10$''$ from \sgra\ in the 2-10 keV band in bins of 180~s) 
from PN data (red crosses).
}
\label{fig1}
\end{figure}

\clearpage
\newpage
\begin{figure}
\plottwo{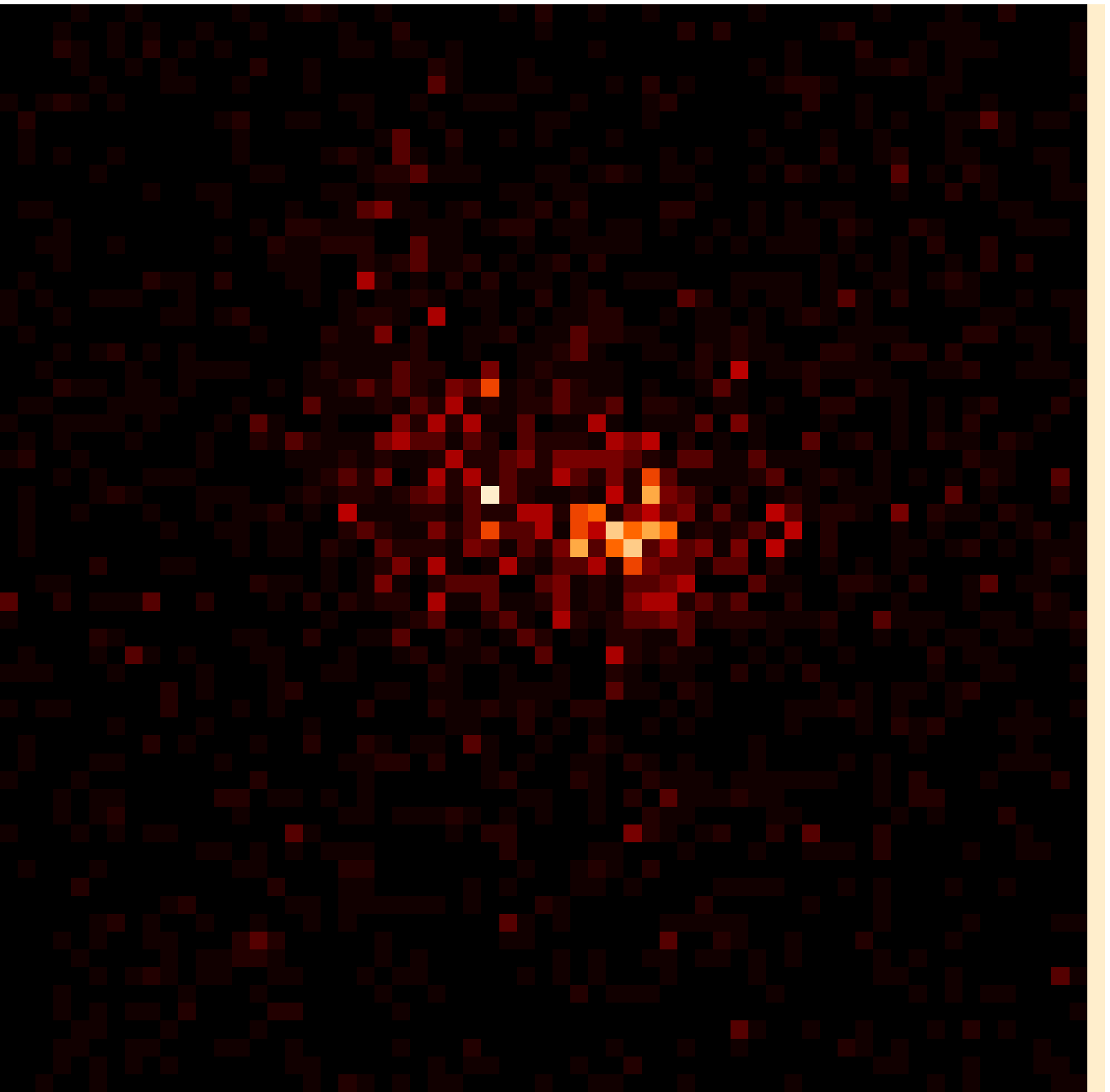}{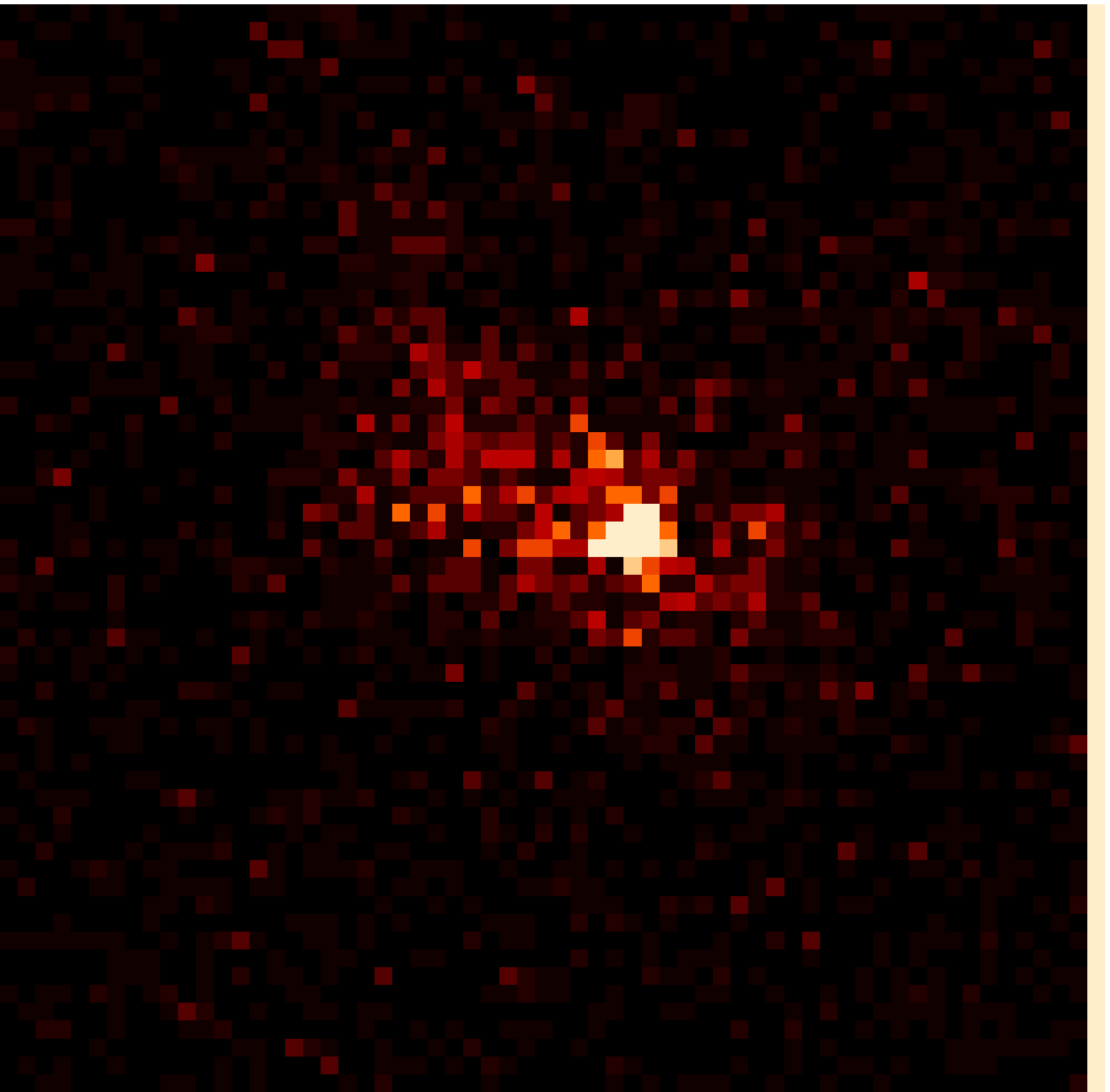}
\caption{Images of the 5$'$~$\times$~5$'$ region around the galactic
nucleus in the band 2-10 keV obtained from MOS events integrated
in the 1000~s before the flare (left) and in the last 1000~s of the 
observation including the flare (right). 
Pixels were rebinned to a size of 5.5$''$~$\times$~5.5$''$. 
\sgra\ position is right in the middle of the central bright pixel
visible in the flare image (right).
}
\label{fig2}
\end{figure}

\clearpage
\newpage
\begin{figure}
\plotone{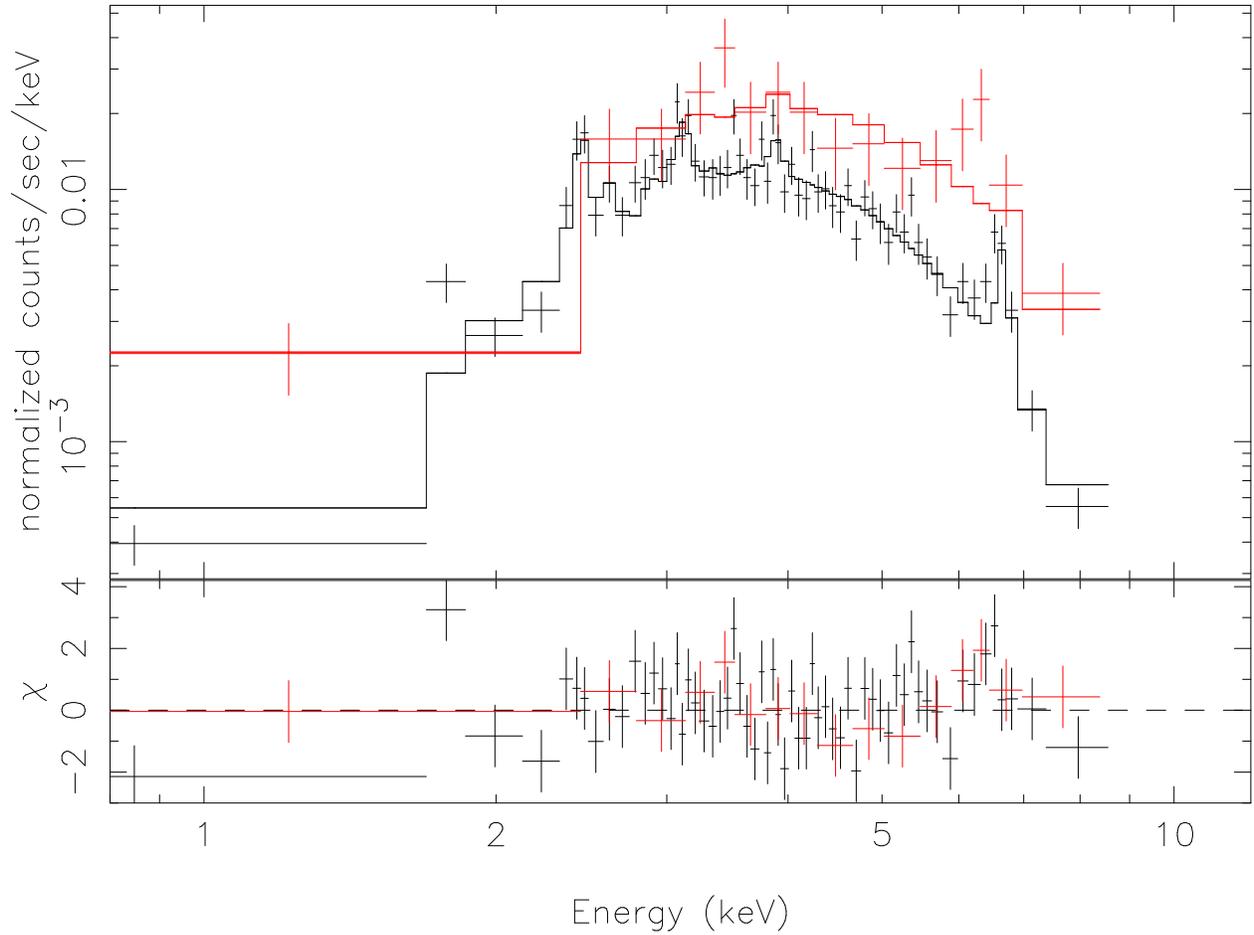}
\caption{MOS count spectra extracted from a region of 10$''$ radius
around \sgra\ before the flare (black, lower data point set)
and during the 900~s flare (red, upper data point set)
compared to the best fit models of Table~1.
}
\label{fig3}
\end{figure}

\clearpage
\newpage
\begin{figure}
\plotone{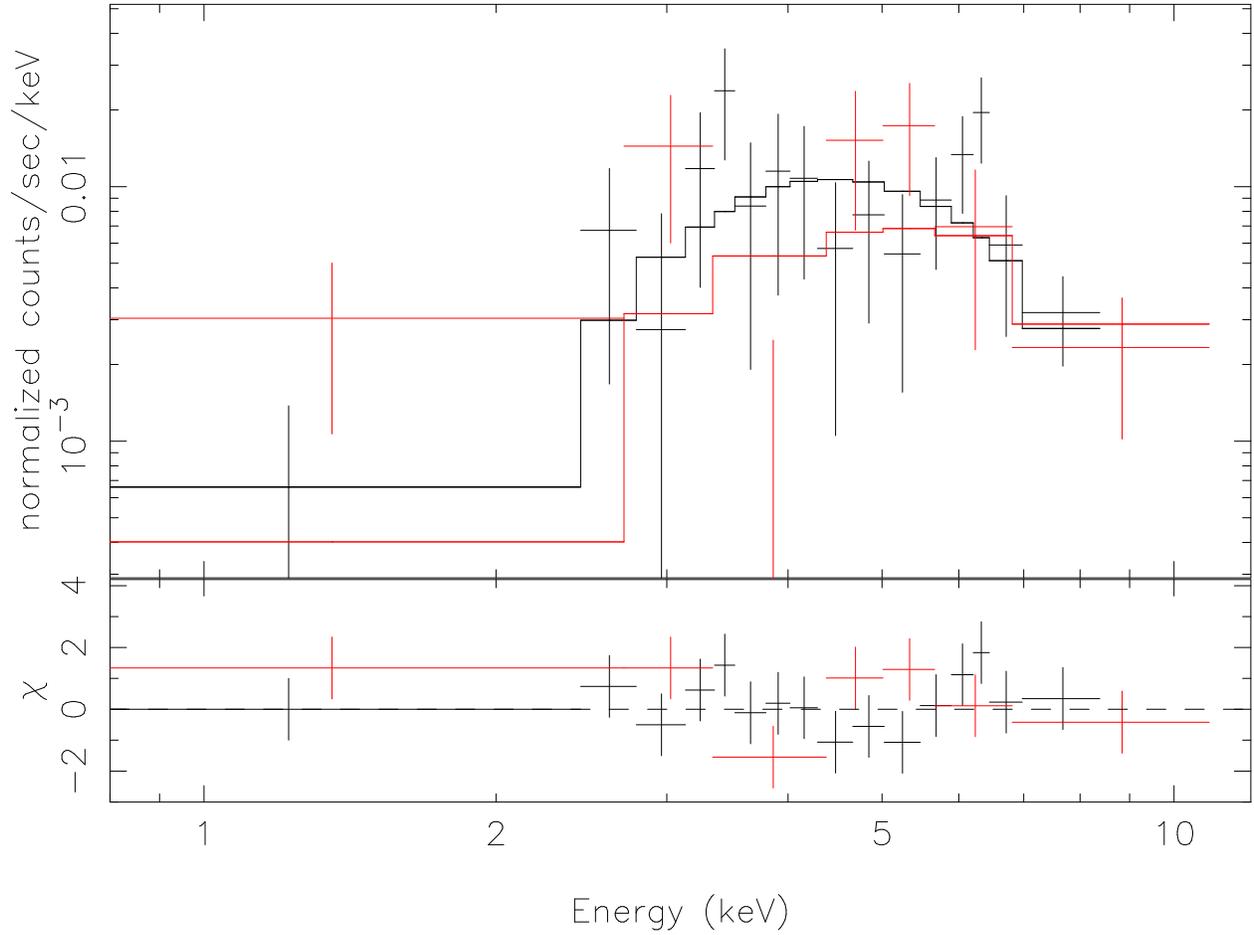}
\caption{Count spectra from MOS (black data point set) and 
PN (red data point set) data, extracted from a region of 10$''$ radius
around \sgra\ during the flare after subtraction of the non flaring 
spectra. 
The spectra are compared to the best fit model of an absorbed power law 
without dust scattering (see Table 2). 
}
\label{fig4}
\end{figure}

\end{document}